# Quantum lithography by coherent control of classical light pulses

Avi Pe'er, Barak Dayan, Marija Vucelja, Yaron Silberberg and Asher A. Friesem

*Department of Physics of Complex Systems, Weizmann Institute of Science, Rehovot 76100, Israel*
*avi.peer@weizmann.ac.il*

*http://www.opticsexpress.org*

**Abstract:** The smallest spot in optical lithography and microscopy is generally limited by diffraction. Quantum lithography, which utilizes interference between groups of *N* entangled photons, was recently proposed to beat the diffraction limit by a factor *N*. Here we propose a simple method to obtain *N* photons interference with classical pulses that excite a narrow multiphoton transition, thus shifting the "quantum weight" from the electromagnetic field to the lithographic material. We show how a practical complete lithographic scheme can be developed and demonstrate the underlying principles experimentally by two-photon interference in atomic Rubidium, to obtain focal spots that beat the diffraction limit by a factor of 2.

## References and links

1. T. A. Brunner, "Why optical lithography will live forever", J. Vac. Sci. Technol. B **21**, 2632-2637 (2003).
2. J. W. Goodman, *Introduction to Fourier optics*, 3rd Ed., (McGraw-Hill, 1996).
3. S. Kawata, H. Sun, T. Tanaka and K. Takada, "Finer features for functional microdevices ", Nature **412**, 697-698 (2001).
4. A. N. Boto, P. Kok, D. S. Abrams, S. L. Braunstein, C. P. Williams and J. P. Dowling., " Quantum Interferometric Optical Lithography: Exploiting Entanglement to Beat the Diffraction Limit", Phys. Rev. Lett. **85**, 2733-2736 (2000).
5. M. D'Angelo, M. V. Chekhova and Y. Shih, "Two-photon diffraction and quantum lithography", Phys. Rev. Lett. **87**, 013602 (2001).
6. K. Edamatsu, R. Shimizu and T. Itoh, "Measurement of the photonic de Broglie wavelength of entangled photon pairs generated by spontaneous parametric down-conversion", Phys. Rev. Lett. **89**, 213601 (2002).
7. B. Dayan, A. Pe'er, A. A. Friesem, and Y. Silberberg, "Nonlinear interactions with an ultrahigh flux of broadband entangled photons", quant-ph/ p. 0411023 (2004), http://xxx.lanl.gov/abs/quant-ph/0411023
8. V. Blanchet, C. Nicole, M. A. Bouchene and B. Girard, "Temporal Coherent Control in Two-Photon Transition: From Optical Interferences to Quantum Interferences", Phys. Rev. Lett. **78**, 2716-2719 (2002).
9. D. Meshulach and Y. Silberberg, "Coherent quantum control of two-photon transitions by a femtosecond laser pulse", Nature **396**, 239 (1998).
10. S.A. Hosseini and D. Goswami, "Coherent control of multiphoton transitions with femtosecond pulse shaping", Phys. Rev. A **64**, 033410 (2001).
11. N. Dudovich, T. Polack, A. Pe'er and Y. Silberberg, "Coherent control with real optical fields: A simple route to strong field control", Submitted to Phys. Rev. Lett (2004).
12. E. Yablonovitch and R. B. Vrijen, "Optical projection lithography at half the Rayleigh resolution limit by two-photon exposure", Opt. Eng. **38**, 334-338 (1999).
13. D. V. Korobkin and E. Yablonovitch, "Twofold spatial resolution enhancement by two-photon exposure of photographic film", Opt. Eng. **41**, 1729-1732 (2002).
14. N. Dudovich, B. Dayan, S. M. Gallagher Faeder, and Y. Silberberg, "Transform-Limited Pulses are not Optimal For Resonant Multiphoton Transitions", Phys. Rev. Lett. **86**, 47-50 (2001).
15. B. Dayan, A. Pe'er, A. A. Friesem, and Y. Silberberg, "Two-photon absorption and coherent control with broadband down-converted light", Phys. Rev. Lett. **93**, 023005 (2004).

Over the last several decades, optical lithography has evolved into a major technique of the semiconductor industry (and elsewhere) for writing fine features on a surface. Due to the high demand for miniaturization, great effort has been invested in obtaining the highest resolution possible in practical lithographic schemes [1]. The minimal feature size that can be written / imaged in linear optics lithography is limited by diffraction to $d_1 \sim \lambda/NA$, where $\lambda$ is the optical wavelength and *NA* is the numerical aperture of the imaging setup [2]. Essentially, the minimal spot is generated by the interference between many plane waves arriving to the substrate at different angles (as high as allowed by the numerical aperture).

The use of non-linear *N*-photon absorption in the lithographic medium leads to an improvement of factor $\sqrt{N}$ in the spot size ($d_N = d_1/\sqrt{N}$) due to the enhanced contrast of the material response, as was indeed demonstrated [3]. Further improvement of *N*-fold over the diffraction limit was suggested by use of "Quantum Lithography" [4-6]. Unlike standard optical lithography, where the single photon acts as the interfering entity, in quantum lithography the interference is between groups of *N* entangled photons that arrive at different angles. This multi-photon interference is equivalent to single-photon interference with a wavelength of $\lambda/N$, leading to the desired improvement in resolution. In order to obtain a spot size $d_1/N$ with quantum lithography, two conditions must be met: First, the absorption in the lithographic material should be purely *N*-photonic. Second, the photons should interfere only as *N*-photons entangled groups, but not as single photons; i.e. the absorption of the *N* photons should be either from one direction or another, excluding possibilities of absorbing some photons from one direction and the rest from another [5,7]. This apparently implies the need for either time-energy or momentum-space entanglement. Quantum entanglement is not easily generated or maintained. Specifically, a severe inherent power limit is imposed if the light is to be considered as a flux of entangled photons (intuitively, the flux must be low enough that the entangled *N*-photon groups arrive "one at a time") [7]. This need to operate at the single-photon regime severely restricts the practicality of quantum lithography.

In this paper we demonstrate that field quantization effects are not necessary and can be replaced by the quantum nature of the lithographic material (i.e. a spectrally narrow *N*-photons transition). As a result, higher powers can be utilized according to standard practical constraints without any inherent limitation. The main principle in our approach is that when a short pulse excites a narrow transition in the material, the excitation lifetime is much longer than the pulse duration. Thus, if the transition is excited again by another pulse within the excitation lifetime of the first pulse, these two excitations can interfere even if the two pulses do not (i.e. are mutually incoherent). Since this interference occurs through the medium, the relative phase that affects it is dictated by the transition frequency $\omega_A$ and the relative delay $\tau$ between the pulses ($\phi = \omega_A \tau$). If this excitation is non-linear of order *N*, the center frequency of the exciting pulse is $\omega_0 = \omega_A/N$. As a result, this "quantum interference" is equivalent to one-photon interference with a wavelength shorter by a factor of *N*, as was indeed demonstrated for *N*=2 [8]. This long lived interference through the medium, between very short pulses is the classical equivalent of time-energy entanglement in quantum lithography; the narrow bandwidth of the transition imposes *N*-photon coherence between mutually incoherent (non-overlapping) pulses. In the following we theoretically analyze and experimentally demonstrate how control over the spatial dependence of the "quantum phase" $\phi(r) = N\omega_0 \tau(r)$ leads to a complete lithographic scheme with a resolution that is improved by a factor of *N*.

A basic question in lithography is how to generate the smallest possible spot with a given lithographic lens. We wish to answer this question for our quantum lithography scheme. The analysis given here assumes one transverse dimension for simplicity (a generalization to two dimensions is straightforward). In standard lithography a small spot is achieved simply by focusing a plane wave with a lens, generating a spot of size $d_1 \approx \lambda f/D$, where f is the focal length of the lens, D is the beam diameter and paraxial optics is assumed. We wish to obtain an additional factor of $N$ in resolution using quantum interference of a train of pulses. Mathematically, the interfering entity is the amplitude of the non-linear response in the lithographic material. In the absence of intermediate resonant transitions this amplitude is proportional to the resonant frequency component of the $N$th power of the electric field $E^N(\omega_A)$ [9,10]. Assuming for simplicity two exciting pulses ($E_1$ and $E_2$), the intensity of excitation will be

$$I(x) \propto \left| E_1^N(x;\omega_A) + E_2^N(x;\omega_A) \right|^2, \quad (1)$$

where $x$ is the spatial coordinate. Note that since the pulses do not overlap in time, mixed terms (e.g. $E_1^p E_2^{N-p}$) are absent. While this expression assumes a weak field perturbative regime, extension to stronger fields is possible [10,11] without change to the relevant features.

The key element in achieving a small spot is to shape the spatial phase fronts of the exciting pulses at the focus such that they will interfere constructively in the center of the spot and destructively near the edges. Specifically, we are searching for $M$ pulse fields at the lens surface $\{\varepsilon_k(x_l)\}$ whose corresponding focal fields $\{E_k(x_f)\}$ fulfill

$$\left| \sum_{k=1}^{M} E_k^N(x_f) \right|^2 = I(x_f), \quad (2)$$

where $I(x_f)$ is the desired narrow lithographic spot. Two constraints must be considered. First, the focal spots cannot be smaller than the diffraction limit imposed by the aperture of the lens. Second, due to the non-linearity, spreading the energy either in space or in time is not desired, so the number of pulses $M$ and the spatial extent of the focal fields should be minimized (i.e. maximize the spatial extent at the lens). While a general solution is currently unknown to us, a simple practical approach is to divide the lens into two non overlapping segments and delay the pulse in one of the segments (e.g. with a piece of glass) as schematically sketched in Fig. 1. Since each segment can be considered as an off axis lens, the two segments generate overlapping spots with linear phase fronts of opposite slopes. If the delay between the pulses is tuned correctly, this will lead to the desired constructive interference at the center. It is interesting to note that a Fourier equivalent of the segmentation scheme of Fig. 1 was suggested for doubling the resolution with two-photon absorption [12,13], where different *spectral* components are separated spatially to two segments on the lens. While such an approach can yield similar results for the two-photon case, generalization to the $N$ photon case is not straight forward and its implementation is more complicated.

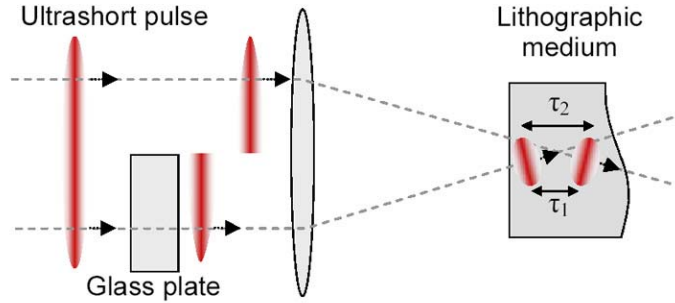

Fig. 1: Schematic setup for generation of sub diffraction limited spots by quantum interference. A glass plate delays half of a planar ultrashort pulse with respect to the other half. As a result, the non-linear lithographic medium at the focus is excited by two consecutive pulses with a space-variant relative delay; thus generating a space-dependent two-photon interference. Fine tuning the delay can be performed by a small tilt of the glass.

Using Eq. (2) we calculated the sub-diffraction limited spots for the arrangement Fig. 1, as compared to the single-photon diffraction limited spot assuming paraxial optics. The results are presented in Fig. 2. Fig. 2(a) and 2(b) show the resulting focal intensity spots for two-photon absorption and four-photon absorption. As evident, the higher resolution is indeed achieved, but at the cost of side lobes, which increase for higher order non-linearity. The side lobes are due to the fact that constructive interference does not occur only at the center, but also at the very edges of the spot. Dividing the lens into more segments [12] only moves the side lobes farther away from the main peak without significant attenuation, as can be seen in Fig. 2(c) and 2(d). The reason for this is that when the lens is divided into smaller segments, each segment leads to a larger one-photon spot. Thus, although the side lobes close to the main peak are attenuated, new side lobes appear at the edges of the larger one-photon spot. Moreover, as the number of segments increase, the energy of the incident pulse spreads in both space and time, leading to an undesired decrease in the material response by a factor of $M^{2(N-1)}$, where $M$ is the number of segments (pulses). While the optimal spatio-temporal distribution for this task is not known, a practical solution is to add only one more pulse to suppress the side lobes. Since the side lobes are out of phase with the main lobe, each side lobe can be cancelled by a pulse that would focus onto the side lobe with appropriate phase. Since the side lobes are far enough apart, only one pulse with two regular foci (i.e. composed of two tilted plane waves) is needed, as was done to obtain the results of Fig. 2(e) and 2(f).

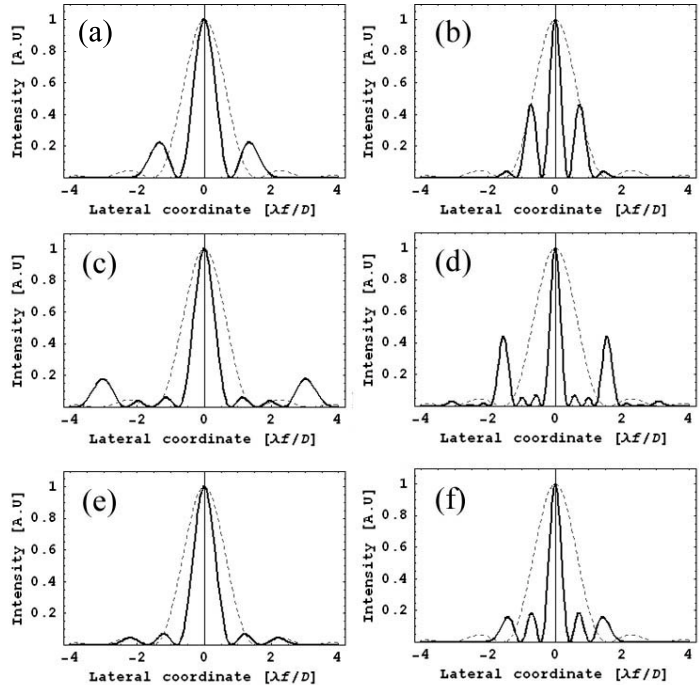

Fig. 2: Calculated sub diffraction-limit spots (line), as compared to the diffraction-limited single-photon spot (dashed). (a), (b) are the two-photon and four-photon spots respectively of the two segment configuration of Fig. 1. (c), (d) show the same spots assuming four equal non-overlapping segments instead of two. (e), (f) show these spots when a third pulse with two offset foci is used to suppress the side lobes. All segments are assumed to be illuminated by a uniform plane wave pulse.

Once a single small spot is achieved, it is clear that any pattern can be written on a substrate by scanning the spot over it. Yet, Unlike linear optical lithography, an arbitrary pattern cannot be considered as a linear superposition of single spots, since mixing can occur between close, but non overlapping, spots. Even though the *N*-photon spots do not overlap, mixing occurs because the larger one-photon spots that generate them do, and interfere. Thus, our scheme is not directly suitable for imaging lithography of arbitrary patterns. However, any use of non-linearity inherently favors scanning because of power considerations. Note that even though the intensity of the non-linear lithographic response with entangled photons depends *linearly* on the incoming photon flux, the above power considerations also apply, since the sensitivity to spatial expansion remains non-linear [5,7].

We chose to demonstrate the principles of our quantum lithography scheme in general and the ability to generate sub diffraction-limited spots in particular, using a narrowband two-photon transition in atomic Rubidium. Our experimental configuration (shown in Fig. 3), is essentially a realization of the basic setup of Fig. 1. The cylindrical telescope weakly focuses in one dimension pulses (~100fs around 778nm) emitted from a Ti:Sapphire laser into the Rb cell. The delay line induces a variable relative delay between the two halves of the pulse. We detected the two-photon excitation by imaging the resulting fluorescence at 420nm onto an Enhanced CCD camera. Since our measurements were performed only in one dimension, the beam was tightly focused in the perpendicular dimension with a strong cylindrical lens in front of the cell in order to increase the signal. Since diffusion of the atoms in the gas during the fluorescence lifetime may blur the two-photon spot, we added a buffer gas (Neon at a pressure of 400Torr) to the Rubidium cell, thus slowing down the spread of the excited atomic cloud by elastic collisions with the buffer. In addition, we used a rather weak focusing into the cell to generate relatively large spots that were not easily washed out.

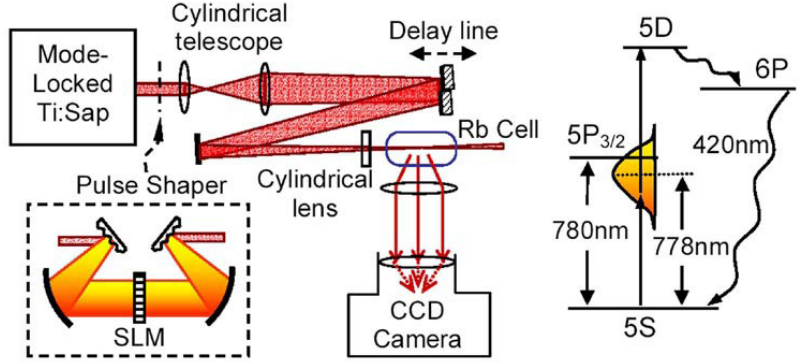

Fig. 3: Experimental configuration and relevant level diagram for atomic Rb. The cylindrical telescope weakly focuses the beam into the Rb Cell, the delay line controls the interference and the CCD records the image of the fluorescence spot. The cylindrical lens in front of the cell tightly focuses the beam in the perpendicular dimension.

We chose for our experiment the 5S-5D two-photon transition centered at 778nm (see the relevant level diagram in Fig. 3). While this transition is relatively strong, we had to avoid the excitation of the intermediate 5P level (at 780), since the resonant two-step excitation via that level has a very long lifetime that would have created undesired interference between the pulses, even at very long delays. Thus, we blocked the resonant frequency at 780nm in a pulse shaper that was placed at the entrance to the experimental configuration. We also utilized the pulse-shaper to introduce a $\pi$ phase shift to frequencies above the resonance and below it, in order to maximize the excitation [14]. Note, that in a normal situation with a non-resonant two-photon transition, the pulse shaper is not required at all.

The experimental results are shown in Fig. 4. The double resolution is demonstrated clearly in Fig. 4a, where two CCD images and the corresponding transverse line cross sections (along with theoretical fits) are presented. The images were taken when the delay line was tuned to obtain a "dark spot" (i.e. destructive interference in the center of the spot). At a delay shorter than the coherence length of the pulse, regular one-photon interference is observed. Yet, when the delay was tuned far beyond the coherence length, two-photon interference is observed and the distance between the lobes is reduced to a half. Fig. 4(b) shows a "bright spot" (constructive at the center) at the two-photon interference regime, demonstrating the two-fold narrowing of the central lobe, as well as the expected side lobes. The discrepancy between the theoretical and experimental results in Fig. 4 is mainly due to imperfect blocking of the resonant component at 780nm in our shaper that led to a residual narrowband response, and partly due to vibrations of the delay line mirrors during the CCD capture time and to diffusion of the Rb cloud during fluorescence.

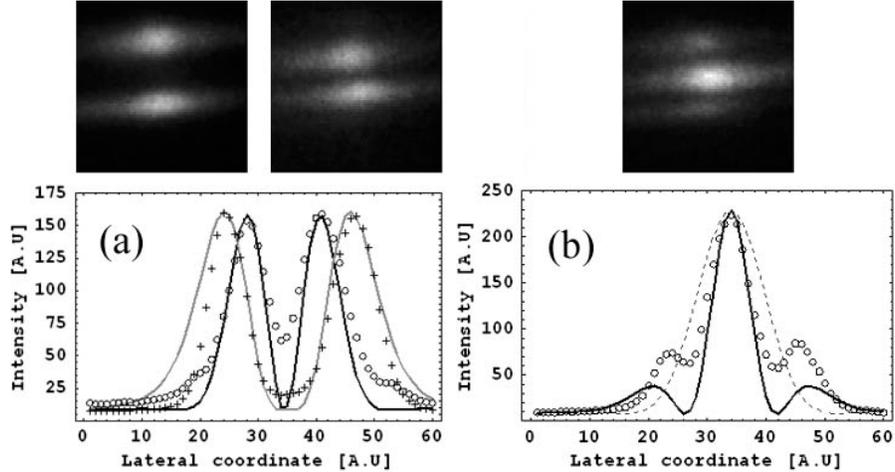

Fig. 4: Experimental results. (a) images and transverse cross sections of "dark spots" (destructive at the center) for a short relative delay (crosses – data, gray line – theoretical fit) and a long relative delay (circles – data, line - theoretical fit), demonstrating the double resolution of two-photon interference compared to one-photon interference. (b) is the corresponding two-photon "bright spot" as compared to the diffraction limited one-photon spot (dashed). All experimental cross sections were averaged over the center portion of the image (18 pixel lines) to reduce noise. The theoretical fits assume a Gaussian beam profile with a narrow gap in the middle.

It is interesting to note that while our discussion focused on coherent ultrashort pulses, a coherent pulsed excitation is not necessary. Since narrow two-photon transitions are unaffected by anti-symmetric spectral phase functions [9], broadband high-power down-converted light, which is both continuous and incoherent can be the source of illumination just as well because of it's inherent anti-symmetric spectral phase [15].

To conclude, we theoretically analyzed and experimentally verified a simple and practical scheme for sub diffraction limit quantum lithography which relies on the quantum nature of the lithographic material and not of the exciting field. In order for the method to be practical, a non-linear lithographic material with a narrow excitation line is required. We speculate that such a material should include some atomic or ionic doping that will trigger the lithographic chemical reaction. We hope that the relatively simple implementation of this method will lead to the development of such materials and to practical applications. While lithography was the major application considered here, the ability to excite sub-diffraction limit spots may be applied to non-linear fluorescence microscopy just as well.